\documentstyle[12pt]{article}
\begin{document}
\thispagestyle{empty}
\begin{flushright}
CUPP-98/3\\
\texttt{hep-ph/9812249} \\
December 1998\\
\end{flushright}
\vskip 35pt

\begin{center}
{\Large {\bf SOLAR NEUTRINO OSCILLATION DIAGNOSTICS
AT SUPERKAMIOKANDE AND SNO}}

\vskip 30pt

{\sf Debasish Majumdar  and Amitava Raychaudhuri}

\vskip 20pt

{\it Department of Physics, University of Calcutta,\\ 
92 Acharya Prafulla Chandra Road, Calcutta 700 009, India} 

\vskip 20pt

{\bf ABSTRACT}

\end{center}

\vskip 10pt

Results for solar neutrino detection from the  SuperKamiokande
collaboration have been presented recently while those from the
Sudbury Neutrino Observatory are expected in the near
future. These experiments are sensitive to the $^8$B neutrinos
from the sun, the shape of whose spectrum is well-known but the
normalisation is less certain. We propose several variables,
insensitive to the absolute flux of the incident beam, which
probe the shape of the observed spectrum and can sensitively
signal neutrino oscillations. They provide methods to extract the
neutrino mixing angle and mass splitting from the data and also
to distinguish oscillation to sequential neutrinos from those to
a sterile neutrino.

\vskip 30pt

\begin{center} 
\end{center}
\newpage

The recent evidence of neutrino oscillations in the atmospheric
neutrino data presented by SuperKamiokande (SK) \cite {ska} has
moved neutrino physics to the centrestage of research activity.
A non-zero neutrino mass will have impact on many areas of
particle physics, astrophysics, and cosmology \cite {numass} and
new results are eagerly awaited. It is widely expected that
important information will emerge from the data on solar
neutrinos.  All earlier experiments have consistently signalled a
depletion of the solar neutrino flux \cite{solar} and high
statistics results from SK (some already published \cite{sks})
and the other experiment of comparable size, the Sudbury Neutrino
Observatory (SNO) \cite{sno}, will further sharpen the situation.

There are several issues pertaining to the solar neutrino problem
which still remain unsettled. The observed flux depletion could
be a consequence of vacuum neutrino oscillations or resonant
flavour conversion \cite{msw}. It is not possible to rule out any
of these alternatives on the basis of the available data.
Further, the electron neutrino may be mixed with either a
sequential or a sterile neutrino. Three neutrinos are expected in
association with the three known charged leptons. The inclusion
of a fourth neutrino -- sterile, in view of the LEP and SLC
results -- is suggested from the several evidences indicative of
neutrino oscillations, namely, the solar neutrino puzzle, the
atmospheric neutrino anomaly and the results of the LSND
experiment, all of which cannot be accommodated together in a
three neutrino framework \cite{goswami}. Finally, it is expected
that the mass splitting and mixing angle will be tightly
constrained from the new data.

The solar neutrinos are produced in standard reactions
(the {\em p-p} chain, CNO cycle, {\em etc.}) responsible for the
generation of heat and light. Though the spectrum of neutrinos
from each of the processes is well known, their absolute
normalisations vary from one solar model to another \cite
{bah,turck}. The two latest detectors, SNO and SK, are sensitive
to neutrinos from only the Boron reaction in the {\em p-p} chain
whose normalisation, for example, is known to vary like
$T_c^{18}$, where $T_c$ is the temperature at the solar core.

In this paper we examine the vacuum oscillation scenario.
We propose several variables relevant for SK and SNO which are
insensitive to the absolute normalisation of the $^8$B neutrino
flux and may be used (a) to distinguish mixing of the electron
neutrino  with a sequential neutrino from that to a sterile
neutrino and (b) to determine the neutrino mass splitting and
mixing angle. Other variables, insensitive to the absolute
normalisation of the incident flux, have been explored earlier in
refs. \cite{blsno,fior} where the focus has been on the energy
spectrum of the scattered electron neutrino at SNO, the MSW
mechanism etc.

The SuperKamiokande detector uses 32 ktons of light water in
which electrons scattered by $\nu_e$ -- through both charged
current (CC) and neutral current (NC) interactions -- are
identified {\em via} their \u{C}erenkov radiation.  If a
sequential neutrino is produced by oscillation, it will
contribute to the signal only through the NC interactions
(roughly one eighth of the $\nu_e$ case) while a sterile neutrino
will be entirely missed by the detector. The SNO detector has
1 kton of $D_2O$ and neutrinos are primarily detected through the
charged and neutral current disintegration of the deuteron: $\nu
+ d \rightarrow e^- + p + p, \,\, \nu + d \rightarrow \nu + p +
n$, respectively.  While the $e^-$ in the CC reaction is
identified through its \u{C}erenkov radiation and can be used to
determine the shape of the incident neutrino spectrum, the NC
measurement, signalled by the detection of the neutron, is
calorimetric.  If oscillations to sequential neutrinos occur then
they will not contribute to the CC signal while the NC channel
will be unaffected.  On the other hand if the $\nu_e$ oscillates
to a sterile neutrino, which has no interactions whatsoever,
then both the CC and NC signals will suffer depletions.

The first class of variables to probe the shape of the observed
neutrino spectra that we propose are $M_n$, the normalised $n$-th
moments of the solar neutrino distributions seen at SK and SNO.
Specifically,
\begin{equation}
M_n = \frac{\int N_{i}(E) E^n dE}{\int N_{i}(E) dE} 
\end{equation}
where $i$ stands for SK or SNO. It is seen from the definition
that the uncertainty in the overall normalisation of the incident
neutrino flux cancels out from $M_n$.

To see how these variables are affected by neutrino oscillations,
first consider oscillation of the electron neutrino to a
sequential neutrino, say $\nu_\mu$.  Since the oscillation
probability is a function of the energy, the shape of the
spectrum will be affected. As noted earlier, at SK the muon
neutrino will only undergo NC reactions. Thus, for
oscillation to a sequential neutrino, we have
\begin{equation}
N_{SK}(E) = \epsilon_{SK} f(E) \left\{P_{\nu_e \rightarrow \nu_e}(E, \Delta,
                             \vartheta) \sigma^{e}_{SK}(E)
                       + P_{\nu_e \rightarrow \nu_{\mu}}(E, \Delta,
                    \vartheta) \sigma^{\mu}_{SK}(E)\right\} N^0_{SK}
\label{NSK}
\end{equation}
Here, $f(E)$ stands for the incident Boron-neutrino fluence,
$\epsilon_{SK}$ for the detection efficiency which, for the sake
of simplicity, is assumed to be energy independent,  and
$N^0_{SK}$ for the number of electrons in the SK detector off
which the neutrinos may scatter. $\sigma^{e}_{SK}(E)$ is the
$\nu_e$ scattering cross-section with both NC and CC
contributions whereas $\sigma^{\mu}_{SK}(E)$ is the $\nu_{\mu}$
cross-section obtained from the NC interaction alone.

Only the CC contributions are relevant at SNO for the determination of
the spectrum and we get:
\begin{equation}
N^{c.c}_{SNO}(E) = \epsilon^{c.c.}_{SNO} f(E) P_{\nu_e \rightarrow
\nu_e}(E, \Delta,\vartheta)\sigma^{c.c.}_{SNO}(E) N^0_{SNO}
\label{NSNOcc}
\end{equation}
$N^0_{SNO}$ is the number of deuteron nuclei in the SNO detector
and $\epsilon^{c.c}_{SNO}$ represents the CC detection efficiency
assumed to be independent of the energy.

If the $\nu_e$ oscillates to a sterile neutrino, which is
decoupled from the weak interactions, it will escape the SK and
SNO detectors completely.  Thus, for sterile neutrinos, in place
of eq. (\ref{NSK}) we have
\begin{equation}
N_{SK}(E) = \epsilon_{SK} f(E) \left\{P_{\nu_e \rightarrow
\nu_e}(E, \Delta,\vartheta) \sigma^{e}_{SK}(E)\right\} N^0_{SK}
\end{equation}
while eq. (\ref{NSNOcc}) is unchanged.

In the two-flavour case, the probability of an electron neutrino
of energy $E_{\nu}$ to oscillate to another neutrino 
(sequential or sterile), $\nu_x$, after the traversal of a
distance $L$ is:
\begin{equation}
P_{\nu_e \rightarrow \nu_x} = \sin^2(2\vartheta) \sin^2 \left
(\frac {\pi L} {\lambda} \right )
\label{ex}
\end{equation}
where $\vartheta$ is the mixing angle, and the oscillation length,
$\lambda $, is given in terms of the mass-squared difference
$\Delta$ by:
\begin{equation}
\lambda = 2.47 \left (\frac
{E_\nu} {\rm MeV} \right ) \left (\frac {\rm eV^2} {\Delta}\right
)\; {\rm metre}
\end{equation}
From probability conservation: $P_{\nu_e \rightarrow \nu_e} =  1
- P_{\nu_e \rightarrow \nu_x}$. 

In Fig. 1 we present the results for $M_1$, $M_2$, and $M_3$ as a
function of the mass splitting $\Delta$ for oscillation to
sequential  as well as sterile neutrinos. Results for the mixing
angle $\vartheta = 45^o$ and 15$^o$ are shown. As expected,
for the smaller mixing angle the effects of neutrino oscillation
are not very prominent. On the other hand, for $\vartheta =
45^o$, the impact of neutrino oscillation is quite significant,
especially for the smaller values of $\Delta$, and it holds
promise for distinguishing between the sequential and sterile
neutrino alternatives.

In order to evaluate the usefulness of these variables in
conjunction with the actual data, it needs to be noted first that
for both the SNO CC and SK signals, what is experimentally
measured {\em via} the \u{C}erenkov technique is the energy of
the outgoing electron. In the case of SNO, the large mass of the
deutreon forces the electron to move in the direction of the
incident neutrino. Further, since the recoiling hadrons are
heavy, the electron's energy equals the incident neutrino energy
less the threshold energy for the CC reaction, 1.44 MeV. For SK
there is a unique correlation between the electron's energy and
scattering angle with the neutrino energy. Thus the neutrino
spectrum can be readily reconstructed from the measured electron
energy for both experiments using the well-known cross-sections
for the appropriate scattering process. The huge sizes of both
detectors ensure that the error in the final results will be
dominated by systematic uncertainties and careful estimates put
these down to a few per cent \cite{blsno}. 

If the errors on the extracted neutrino spectrum are at the
expected few per cent level, it is easy to convince oneself from
Fig. 1 that $M_1$, $M_2$, and $M_3$ will be useful diagnostic
tools. This gives us confidence that, if the mixing angle
$\vartheta$ is not small (as indicated by the data from the other
earlier experiments), the experimental results will enable a
distinction between the sequential and the sterile neutrino
alternatives and help focus on the mixng angle $\vartheta$ and
mass splitting $\Delta$ involved.

We have also considered the ratios of the moments $r_i =
(M_i)_{SK}/(M_i)_{SNO}$ as variables for the search for neutrino
oscillations. We do not discuss these in this preliminary
communication and results will be reported elsewhere \cite{mr}.

The SNO experiment will enable separate measurements of the
neutrino flux through charge current and neutral current
reactions. As noted earlier, if $\nu_\mu$ or $\nu_\tau$ are
produced through oscillation of solar neutrinos then  they will
register {\em via} neutral current interactions with full
strength but their energy will not permit charged current
interactions.  The ratio, $R_{SNO}$, of the calorimetrically
measured signal in the NC channel, $\int N_{SNO}^{n.c.}$, to the
total (energy integrated) signal in the CC channel, $\int
N_{SNO}^{c.c.}$, is therefore a good probe for oscillations. Thus
\begin{equation}
R_{SNO} = \frac{\int N_{SNO}^{n.c.}}{\int N_{SNO}^{c.c.}}
\label{rsno}
\end{equation}
where
\begin{equation}
\int N_{SNO}^{n.c.} = \int \epsilon^{n.c.}_{SNO} f(E)
\sigma^{n.c.}_{SNO}(E) N^0_{SNO} dE
\label{snonc}
\end{equation}
where $\epsilon^{n.c.}_{SNO}$ is the efficiency of detection of
for the NC channel and
\begin{equation}
\int N_{SNO}^{c.c.} = \int \epsilon^{c.c.}_{SNO} f(E) P_{\nu_e
\rightarrow \nu_e}(E, \Delta,
\vartheta)\sigma^{c.c.}_{SNO}(E) N^0_{SNO} dE
\label{snocc}
\end{equation}
Clearly, $R_{SNO}$ is independent of the absolute normalisation
of the incident neutrino flux $f(E)$ and only depends on its shape.

If oscillations to sterile neutrinos take place then  eq.
(\ref{snonc}) is replaced by:
\begin{equation}
\int N_{SNO}^{n.c.} = \int \epsilon^{n.c.}_{SNO} f(E) P_{\nu_e
\rightarrow \nu_e}(E, \Delta,
\vartheta) \sigma^{n.c.}_{SNO}(E) N^0_{SNO} dE
\end{equation}
while eq. (\ref{snocc}) is unchanged.

Results for $R_{SNO}$ are presented in Table 1. For simplicity,
we have assumed $\epsilon^{n.c.}_{SNO}$ to be independent of the
energy and further equal to the efficiency of the CC reaction
$\epsilon^{c.c.}_{SNO}$. If instead,
$\epsilon^{n.c.}_{SNO}/\epsilon^{c.c.}_{SNO} = r_{\epsilon}$ and
it can be taken to be independent of the energy to a good
approximation, then our results for $R_{SNO}$ will be multiplied
by this factor.

If no oscillations take place then we find $R_{SNO}$ = 0.382.
Oscillation to sequential neutrinos decreases the denominator of
eq. (\ref{rsno}) while the numerator is unaffected. Thus
$R_{SNO}$ increases if such oscillations take place.  From Table
1 it is seen that, especially for larger mixing angles $\vartheta
= 30^o$ or $45^o$, $R_{SNO}$ is significantly different from the
no-oscillation limit for the sequential
neutrino case. For the sterile neutrino alternative, the change in
$R_{SNO}$ is very marginal and it is unlikely that it will be
observable. Thus $R_{SNO}$ offers a clear method for the
distinction between the sequential and sterile neutrino
alternatives, independent of the uncertainty in the overall
normalisation of the incident neutrino flux.

In Fig. 2 we present contours of constant values of
$R_{SNO}$ in the $\Delta$-$\vartheta$ plane for oscillation to
sequential neutrinos. The symmetry of the
contours about $\vartheta = 45^o$ is expected. At $\Delta = 0$ or
$\vartheta = 0^o$ or $90^o$ the limit of no oscillations will be
obtained.  Values of $R_{SNO}$ as high as 0.99 can only be
achieved for smaller values of $\Delta$.

In this work, we have considered the oscillation of $\nu_e$ to
either (a) a sequential neutrino or (a) sterile neutrino. We have
restricted ourselves to vacuum neutrino oscillations. We plan to
examine  the alternative of matter enhanced MSW resonant flavour
conversion later.  We have not extended the analysis to a three
(or four) neutrino mixing scheme. This would have introduced too
many parameters.  We have also ignored a small contribution from
{\em hep}-neutrinos. These variables can also be utilised to
study oscillation of supernova neutrinos.  Some results are
presented in ref. \cite{mkrrs}.

The behaviour of the variables $M_1$, $M_2$, and $M_3$ and that of
$R_{SNO}$ leads us to believe that, as data from SuperKamiokande
and SNO accummulate, used in conjunction they may be fruitful
not only to look for oscillations of solar neutrinos but also to
zero in on the mass splitting and mixing angles for solar
neutrino oscillations.

\noindent {\large{\bf {Acknowledgements}}}

This work is partially supported by the Eastern Centre for
Research in Astrophysics. A.R. also acknowledges a
research grant from the Council of Scientific and Industrial
Research.

\newpage
\begin{center}
{\bf TABLE CAPTION}
\end{center}

\noindent Table 1: $R_{SNO}$ for different values of the mixing angle,
$\vartheta$, and the mass splitting, $\Delta$. Results are
presented for both mixing with sequential and sterile neutrinos.

\newpage

\begin{center}
{\bf{\Large Figure Captions}}
\end{center}
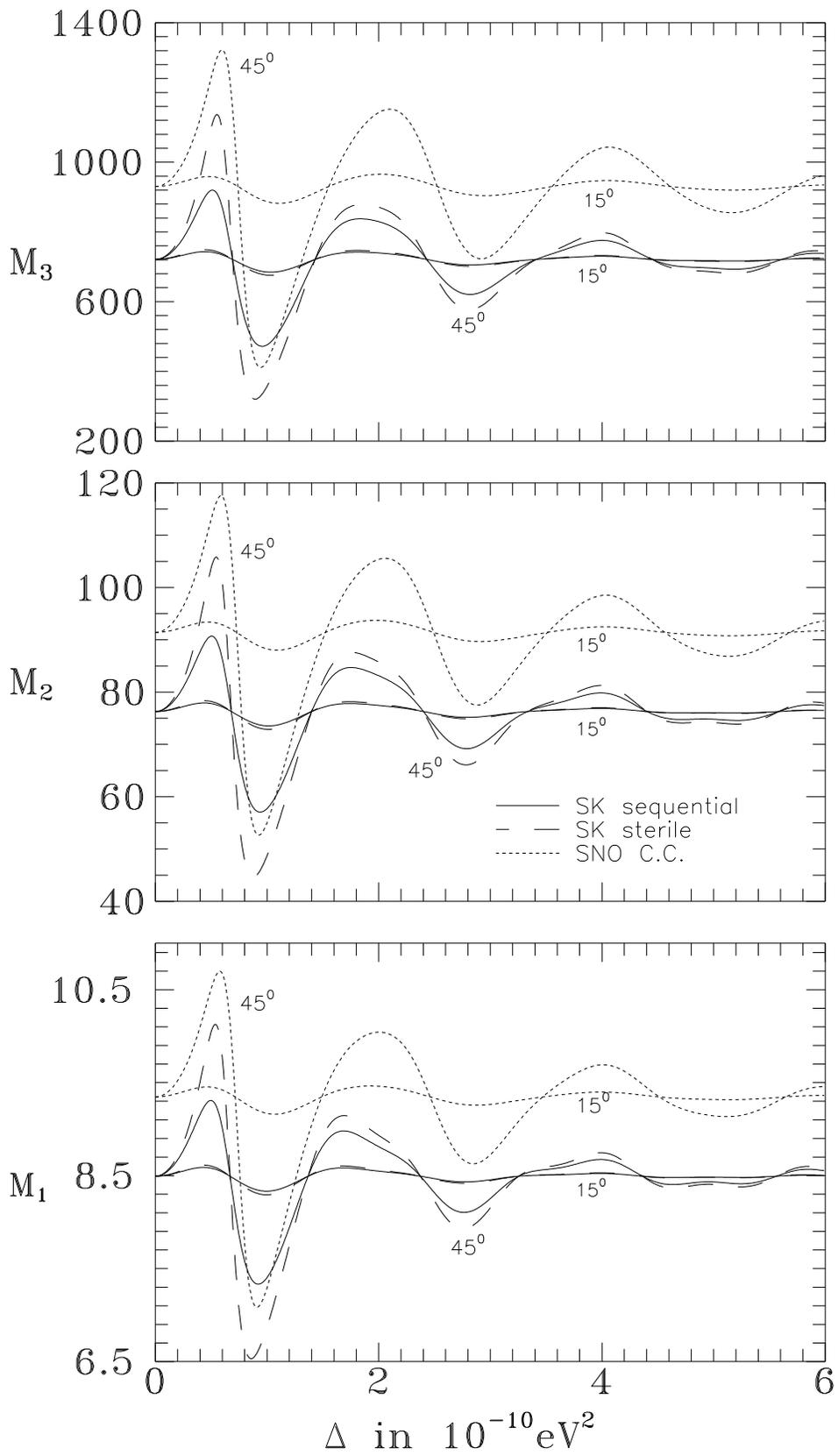
Fig. 1: The variables (a) $M_1$, (b) $M_2$, and (c) $M_3$ as a
function of the mass splitting $\Delta$ for the SuperKamiokande
and SNO detectors. Results are presented for two values (45$^o$ and
15$^o$) of the mixing angle $\vartheta$. Note that the SNO
(charged current) signal does not distinguish between the
sequential and sterile neutrino scenarios. \\
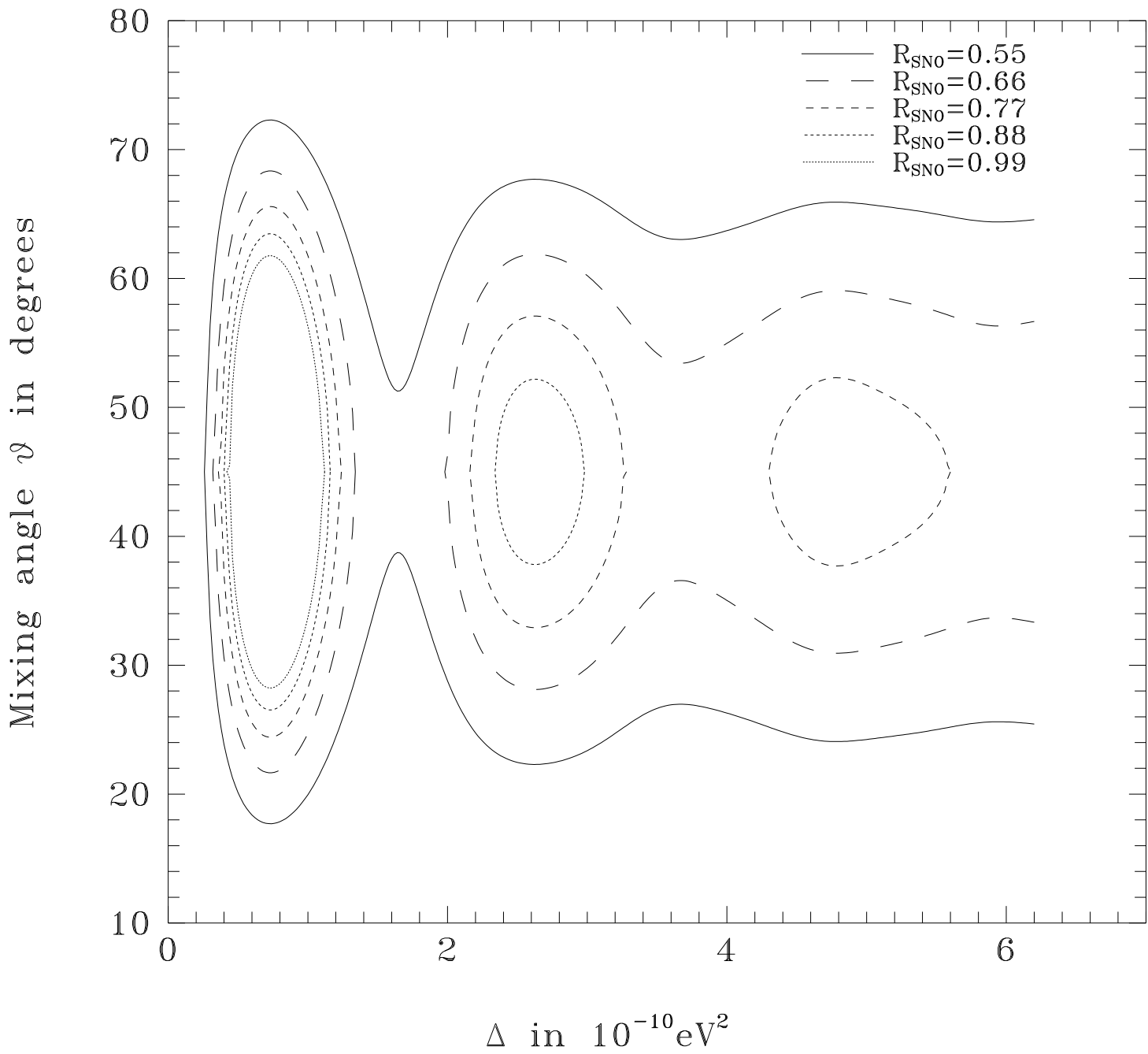
Fig. 2: Contours of constant $R_{SNO}$ -- the ratio of the NC
signal to the energy integrated CC signal at SNO -- in the
$\Delta - \vartheta$ plane for oscillation to sequential 
neutrinos.  No neutrino oscillation corresponds to
$R_{SNO}$ = 0.382.  \\

\newpage
\begin{center}
\begin{tabular}{|c|c|c|c|c|c|c|}
\hline
$\Delta $ & \multicolumn{6}{|c|}{$R_{SNO}$} \\
\cline{2-7}
in & \multicolumn{2}{|c|}{$\vartheta=15^0$} & 
\multicolumn{2}{c|}{$\vartheta=30^0$} & 
\multicolumn{2}{c|}{$\vartheta=45^0$} \\
\cline{2-7} 
$10^{-10}$ eV$^2$& Sequential & Sterile & Sequential & Sterile & Sequential & Sterile \\
\hline

 0.0 &  0.382 &  0.382 &  0.382 &  0.382 & 0.382 &  0.382 \\
 0.3 &  0.422 &  0.384 &  0.532 &  0.389 & 0.613 &  0.392 \\
 0.6 &  0.480 &  0.383 &  0.991 &  0.387 & 2.117 &  0.396 \\
 0.9 &  0.467 &  0.378 &  0.848 &  0.362 & 1.428 &  0.337 \\
 1.2 &  0.438 &  0.380 &  0.623 &  0.375 & 0.788 &  0.370 \\
 1.5 &  0.422 &  0.383 &  0.537 &  0.387 & 0.620 &  0.390 \\
 1.8 &  0.417 &  0.383 &  0.512 &  0.386 & 0.577 &  0.388 \\
 2.1 &  0.431 &  0.383 &  0.582 &  0.387 & 0.706 &  0.390 \\
 2.4 &  0.444 &  0.382 &  0.660 &  0.383 & 0.873 &  0.384 \\
 2.7 &  0.444 &  0.380 &  0.658 &  0.375 & 0.867 &  0.370 \\
 3.0 &  0.444 &  0.381 &  0.659 &  0.379 & 0.869 &  0.377 \\
 3.5 &  0.431 &  0.382 &  0.582 &  0.381 & 0.705 &  0.380 \\
 4.0 &  0.434 &  0.383 &  0.597 &  0.386 & 0.735 &  0.388 \\
 4.5 &  0.435 &  0.382 &  0.606 &  0.381 & 0.753 &  0.380 \\
 5.0 &  0.440 &  0.382 &  0.634 &  0.381 & 0.813 &  0.381 \\
 5.5 &  0.437 &  0.382 &  0.614 &  0.381 & 0.770 &  0.381 \\
 6.0 &  0.434 &  0.382 &  0.597 &  0.382 & 0.735 &  0.382 \\
\hline
\end{tabular}
\end{center}

\end{document}